\documentclass[10pt]{article}
\usepackage{graphicx,color}
\begin{document}
\title{Investigation of correlation effects in FeSe superconductor by LDA+U 
method}
\author{ H. Lohani, P. Mishra and B. R. Sekhar}
\maketitle
\begin{abstract} 
Correlation effects are observed strong  in Iron chalcogenides superconductors 
by experimental and theoretical investigations. We present a comparative study 
of the influence of Coulomb interaction and Hund's coupling in the electronic 
structure of FeSe and FeTe. The calculation is based on density functional 
theory (DFT) with local density approximation(LDA+U) framework employed in 
TB-LMTO ASA code. We found the correlation effects were orbital selective due 
to the strength of interorbital hybridization among different Fe-3d orbitals 
mediated via chalcogen (Se/Te-p) orbitals is different in both the compounds, 
however Coulomb interaction is screened significantly by Te-p bands in FeTe. 
Similarly the orbital section is different in both the compounds because of 
the difference in the chalcogen height.
\end{abstract}
\vspace{2pc}
\maketitle

\section{Introduction}

Iron based superconductors, particularly members of FeSe$_{1-x}$Te$_x$ family 
attract much attention due to their nature of strong electron correlation 
unlike other superconductors. A recent advancement in this field is the 
synthesis of single-layer films of FeSe on SrTiO$_3$ substrates exhibiting 
superconductivity (T$_c$=80 K) which turns insulating with the addition of 
one more layer\cite{Wangg}. This unusual behavioral difference between single 
and double layer films of FeSe is a signature of strong electron correlation 
which has been experimentally observed\cite{Lee,Xu}. Superconductivity in the 
FeSe$_{1-x}$Te$_x$ compounds was first reported by Hsu {\it et al.} \cite{hsu} 
in the FeSe (x = 0) compound exhibiting a T$_c$ around 8 K which rises up to 
37 K under pressure (7GPa)\cite{marg}. On the other hand, the other extreme 
composition of this family, Fe$_{1.068}$Te, though not a superconductor show 
a spin density wave (SDW) ordering at 67K\cite{Dai} with an accompanying 
structural transition from tetragonal to monoclinic. With Se doping 
superconductivity emerges in FeTe with a simultaneous decrease in the SDW and 
the value of T$_c$ reaches maximum 15 K for x = 0.5 doped case. The Fe content 
is also detrimental for superconductivity; with excess Fe favors the spin 
localization destroying the superconductivity\cite{Keller,Marel}. Both FeSe 
and FeTe have tetragonal crystal structure belonging to space group symmetry 
P$_4$/nmm. It consists a square planar sheet of Fe atoms, which is 
tetrahedraly coordinated with anion (Se/Te) atoms. However, the height of 
anion atom from the Fe square plane is different in these two compounds and 
this plays a pivotal role in determining the electronic properties of these 
systems\cite{Kuchinskii,Tou}.

A recent ARPES study on FeSe$_{1-x}$Te$_{x}$ compositions by Ieki {\it et 
al.} \cite{Ieki} has shown clearly the strong electronic correlation in 
these compounds, where a small quasi particle weight in FeTe transforms into 
a sharp one with increase in Se content. Other ARPES results \cite{xia,chen} 
on these compounds have shown significant band renormalization which was 
supported by the calculations based on local density approximation (LDA). 
Tamai {\it et al.} \cite{tamai} has found the mass renormalization factor 
to be m$^*$/m = 20 from their ARPES study on FeSe$_{0.42}$Te$_{0.58}$. It is 
close to the value observed in highly correlated systems like transition metal 
oxides. Also, our angle integrated valence band photoemission study on 
FeSe$_{1-x}$Te$_x$ \cite{mishra} revealed significant spectral weight shifts 
in the near E$_f$ region with Se doping leading to the formation of a 
pseudogap. Further, a temperature dependent orbital selective spectral weight 
transfer was also reported by us\cite{mishra}. Although, such manifestations 
of the strong coulomb correlation were also shown in many other photoemisson 
studies\cite{yoshida,alaska}, these experimental observations are not well 
addressed by LDA based electronic structure calculations. However, results of 
some recent calculations\cite{markus,ansgar,craco}, where Coulomb correlations 
were included by using LDA+DMFT frame work, are very close to the experimental 
findings. In this paper we are presenting our calculations showing the 
evolution of electronic structure by the incorporation of different values of 
Coulomb interaction U and intra-atomic exchange J based on LDA+U scheme in 
FeSe and FeTe. We observed multi orbital correlation effect in Fe-3d states 
which is more prominent in FeSe in comparison to FeTe. We have discussed our 
results referring to the difference in the geometry of the anion tetrahedra 
in both the compounds.

\section{Details of Calculation}

The band structure calculations we performed were based on first principles 
Langreth-Meh-Hu gradient corrected von Barth Hedin parametrized LDA \cite{von} 
energy and potential. Lattice parameters used in the calculations are taken 
from the experimental data\cite{kazumasa,Zarger} published earlier by others. 
Correlation effects of the Fe-3d orbitals have been examined by employing 
different values of Coulomb interaction parameter U and Hund's coupling 
J\cite{vladimir}. Empty spheres were introduced to make the volume of the 
unit cell equal to the total volume of the spheres within the permissible 
limit of atomic sphere approximation (ASA). Fe-4s, 4p, 3d ; Se-4s, 4p, 4d and 
Te-5s, 5p, 5d orbitals are used as the basis set for the valence energy 
region. A mesh of 12$\times$12$\times$8 is used for sampling the irreducible 
part of the Brillouin zone integration. Height of the anion atom from the Fe 
square plane was relaxed by minimizing the total energy using Quantum Espresso 
code\cite{Giannozzi}. We checked our calculation parameters by comparing them 
with the LDA results reported earlier\cite{alaska,grechnev}.

\section{Results and Discussion}
\subsection{DOS}
Fig (1a) and (1b) show the plots of density of states (DOS) for FeSe and FeTe 
respectively for the range of 4.0 to -7.0 eV binding energy (BE). The valence 
band (VB) DOS comprises of the region from E$_f$ to -6.0 eV BE while the 
unoccupied DOS extends from E$_f$ to 3.0 eV BE. The near E$_f$ states from 2.8 
to -2.4 eV are predominantly Fe-3d derived for both FeSe and FeTe. These Fe-3d 
states, separated in lower and upper bands exhibit a clear pseudogap feature 
just above the E$_f$ (0.24 eV) in the case of FeSe whereas it is less 
prominent in the case of FeTe. The states around -2.2 to -6.0 eV are originate 
from the Fe-3d and anion (Se/Te)-p hybridized states. Interestingly, for the 
case of FeSe, there exists a sharp gap at -2.3 eV which is not present in case 
of FeTe. Due to the smaller electronegativity of Te compared to Se, the 
hybridized states between Fe-3d and anion-p orbitals are placed at a lower BE 
in FeTe in comparision to those of FeSe.  DOS have also been calculated after
 downfolding the valence orbital of the anion atom (Fig (1c) and (1d)) 
in order to get a deeper insight into the role of anion orbitals. The gap
 at -2.3 eV present in Fig (1a) is due to the splitting
 of bonding and antibonding bands between Fe-3d and Se-p hybridized states which 
becomes less prominent when the valence orbitals of Se atom are downfolded as 
shown in Fig (1d). Similarly the pseudogap feature across the E$_f$ is also 
associated with the hybridization between Fe-3d and Se-p orbitals. Unlike FeSe, 
the anion orbitals do not play any major role in modifying the DOS in FeTe as 
is clear from Fig (1b) and (1d). It indicates a weak hybridization between 
Fe-3d and Te-p orbitals in FeTe. The role of the anion orbitals is linked to 
the structural geometry of FeSe and FeTe. The insets of Fig (1c) and (1d) show 
the geometry of anion tetrahedra in FeSe and FeTe respectively. This tetrahedral 
geometry depends on two important parameters; firstly, the height of anion 
from the Fe square plane (z) which is 1.64 \AA{} and 1.46 \AA{} in case of 
FeTe and FeSe respectively and secondly, the anion-Fe-anion angle ($\alpha$). 
The enhanced z height in case of FeTe reduces the interorbital hoping among the 
Fe-3d orbitals, mediated via anion p orbitals. Similarly, the value of $\alpha$ 
is 99.9$^\circ$ in case of FeTe which increases to the perfect tetrahedron 
value 109.4$^\circ$ in case of FeSe. The large value of $\alpha$ and small 
value of anion height (z) makes a stronger hybridization between the Fe-3d and 
anion-p orbitals in case of FeSe in comparison to FeTe. This difference in 
hybridization strength is reflected in the plot of DOS Fig (1a) and Fig (1c).

Coulomb correlation effects are important in bands of narrow width, especially 
in Fe-3d states. So the changes in Fe-3d states under the influence of 
different values of U have been calculated and shown for FeSe and FeTe 
respectively in Panel (a) and (b) of Fig 2. In FeSe, Fe-3d states start 
localizing with the application of U and noticeable changes occur at higher 
values of U. For U = 4.0 eV case, the pseudogap feature disappears around 
E$_f$ and only two peaks are observed in the VB comparared to the case of U = 
0.0 eV. These two peaks merge and shift towards higher BE with U = 5.0 eV. In 
case of FeTe, the Fe-3d states also become localized under the application of 
U. However the amount of shift towards higher BE at large values of U (3.5 and 
5.0 eV) is less and states are more at the vicinity of E$_f$ at smaller values 
of U (1.0 and 2.0 eV). In addition to this, narrowing of the lower and upper 
bands of Fe-3d states with increase of U is less. It indicates that the effect of Coulomb 
correlation is weak in FeTe. Possible reason for this, is the presence of Te-p 
states at lower BE which screen the U strongly \cite{takashi}. On the basis of
previous reports \cite{takashi,markus,craco} the value of  U=4.0 and 3.5 eV are 
chosen  to see the evolution of Fe-3d states under the influence of J in
 FeSe and FeTe  as shown in Fig 3(a) and 3(b) respectively. It is observed that
the Fe-3d states are modified significantly even by introduction of a small value 
of J=0.1 eV in case of FeSe. The Hund's coupling, shifts all the Fe-3d states towards
lower BE, with a simultaneous appearance of pseudogap slightly above E$_f$. 
With increase in the value of J further, there is no substantial changes in the DOS.
In the case of FeTe, the Fe-3d states are also shifted towards lower BE, particularly the
states near E$_f$ increase gradually,  with the incorporation of J though the changes are less as compared to FeSe. 
These results show that Hund's coupling J is a key factor in the formation of Fe-3d DOS.\par
In order to heighlight the correlation effects, DOS of different  Fe-3d orbitals are plotted at different values of U and J
in FeSe and FeTe  as shown in Panel (a) and (b) of Fig 4  respectively.
 In FeSe, in the absence of U and J, near E$_f$ states and pseudogap
 feature arises from d$_{yz/xz}$ and d$_{x^2-y^2}$ orbitals. The states originating from 
d$_{xy}$ orbital have largest splitting with two peaks at -1.7 and 1.6 eV BE in the DOS.
Additionally, a clear gap is observed   in d$_{3z^2-r^2}$ states around E$_f$ which are quite 
localized at -0.8 eV. Application of  Coulomb interaction(U=4.0 eV), results in localization of 
the states derived from all four orbitals and the states shift towards higher BE. This shift is 
the largest in the states of d$_{x^2-y^2}$ orbital. Major effect of U is observed  in
d$_{yz/xz}$ states, where broad states near E$_f$ transforms into two clear peaks at higher BE.
Hence the pseudogap feature vanishes across E$_f$. Application of a small value of Hund's coupling
J=0.1 eV, restore the d$_{yz/xz}$ states  near  E$_f$ and no significant changes are seen by
 further increasing the  J value  from 0.1 to 1.2 eV. This nature of pseudogap, which occurs in full range 
of Hund's coupling but absent when J=0.0 at U=4.0 eV, is consistent with the previous work of
 Ansgar {\it et al.} \cite{ansgar} where this pseudogap is attributed to a resonance in self energy caused 
by spin fluctuations. In  case of FeTe in the absence of  U and J values, near E$_f$ states and pseudogap
 are also formed by d$_{yz/xz}$ and d$_{x^2-y^2}$ states like FeSe but a gap is present 
at -0.7 eV in d$_{yz/xz}$ states as well as number of these states are more  across the
E$_f$. The d$_{3z^2-r^2}$ and d$_{x^2-y^2}$ states shift towards  higher BE  by switching on the U=3.5 eV 
and the amount of shift is quite small in comparison to FeSe. These states shows an incremental shift towards E$_f$
 with increase  the J value  in this case. These changes in DOS are presented  in table 1,
 where the occupancy of electrons in different orbitals of Fe-d is tabulated, for  quantitative analysis. 
The occupancy of d$_{yz/xz}$ and d$_{x^2-y^2}$ orbitals show an opposite 
behaviour for FeSe and FeTe under the influence of Coulomb correlation energy. On the other hand,
a remarkable enhancement is observed in the occupancy of d$_{x^2-y^2}$ orbital after introducing 
j= 0.1 eV in FeSe. It is almost twice in comparison to J=0.0 eV case. The change in the
occupancy of Fe-3d orbitals due to the effect of U and J presented here summaries the orbital selective effect in
Fe-3d orbitals in FeSe and FeTe. The individual occupancy of four orbitals are different which are further
enhanced by Coulomb interaction and Hund's coupling. It turns out to a orbital selective nature of the correlation effect which 
is crucially depends on individual band filling factor\cite{Antoine}.\par
\subsection{Band}
Fig 5 shows the near E$_f$ band structure of FeSe with U = 0.0 (5a), U = 4.0 
eV (5b) and FeTe with U = 0.0 (5c) and U = 3.5 eV (5d). 
Intially when U and J = 0.0 eV, three hole like and two electron like bands
are observed at $\Gamma$  and M point respectively in both FeSe and FeTe. However, the
outer hole like bands are quasidegenerate in case of FeSe.
In order to show the contribution of different Fe-3d orbitals, the 
fat bands are calculated for both FeSe and FeTe. These bands are plotted  in Fig 5(e-h) 
for the FeSe case with U and J = 0.0 eV. Fatness of bands 
indicate that the innermost hole like band has a d$_{yz/xz}$ and the outer two 
have d$_{xy}$ and d$_{x^2-y^2}$ orbital characters in both the compounds.
 Similarly the electron like 
bands are composed mainly of d$_{x^2-y^2}$ with little contribution from the 
d$_{yz/xz}$ orbital in FeSe while  inner electron like band (-0.23 eV) has 
d$_{yz/xz}$ and outer one (-0.47 eV) has d$_{x^2-y^2}$ orbital character in FeTe.
Another major difference is that the Te-p bands are intermixed with Fe-d bands 
opposite to that of FeSe, where Fe-d bands are quite separated from the Se-p 
bands. These Te-p bands screen the effect of Coulomb interaction U in FeTe, 
hence the value of U is smaller in FeTe in comparison to FeSe. After applying Coulomb
interaction degeneracy of the d$_{xy}$ and d$_{x^2-y^2}$ hole like 
bands at the $\Gamma$ point has been lifted as shown in Fig(5b). 
The d$_{x^2-y^2}$ band moves towards lower BE and other two d$_{xy}$ and d$_{yz/xz}$
band move towards higher BE at $\Gamma$ point. The same band moves in opposite direction 
in case of FeTe by the application of U= 3.5 eV. On the other hand,
the separation between the two electron like bands at the M point is enhanced in FeTe unlike the FeSe 
case where both the electron like bands are quasidegenerate and shift towards 
the lower BE under the influence of U = 4.0 eV.\par
Evolution of the near Fermi bands in 
FeSe and FeTe, with the tuning of different values of J,
are shown in panel a and b of Fig 6 respectively. In FeSe, at $\Gamma$ point,
 The d$_{x^2-y^2}$ band which was crossing the E$_f$ comes down below the E$_f$ and 
the d$_{yz/xz}$ band which was not crossing E$_f$, crosses E$_f$  after applying  J = 0.1 eV.
Similarly, degeneracy of the two electron like bands at M point is also 
lifted. In FeTe, a gradual shifting is observed in the hole like d$_{x^2-y^2}$ band  
at the $\Gamma$ point towards the E$_f$ and a gradual decrease in the separation between
 two electron like bands at the M point under the influence of Hund's coupling.\par
In a recent ARPES report on FeSe two hole like bands have been observed 
around the $\Gamma$ point and two bands, one electron like and the other hole 
like, at the M point from 40 meV below E$_f$ \cite{Urata}. Another ARPES study 
on FeTe$_{1-x}$Se$_{x}$ for x=0, 0.2, 0.3, 0.4 and 0.45 compounds by Ieki 
{\it et al.} \cite{Ieki} reported three clear hole like bands at the $\Gamma$ 
point ($\alpha$, $\alpha'$ and $\beta$) which evolve with Se doping and a 
shallow electron pocket at the M point for x = 0.45. The $\alpha'$ and $\beta$ 
crosses E$_f$ whereas $\alpha$ is around 20 meV below E$_f$. Similar results 
have also been experimentally observed in Fe$_{1.04}$Te$_{0.66}$Se$_{0.34}$ 
\cite{chen}, FeSe$_{0.42}$Te$_{0.58}$ \cite{tamai}, Fe$_{1.03}$Te$_{0.7}$Se$_{0.3}$ \cite{Rak} 
and FeTe$_{0.55}$Se$_{0.45}$ \cite{Miao}. On the other hand in Fe$_{1.02}$Te 
\cite{Liu} and Fe$_{1-x}$Te/Se \cite{xia} only two hole like bands are 
observed at $\Gamma$ point. The band renormalization factor also varies highly 
at different points of Brillouin zone. For example, in case of 
FeSe$_{0.42}$Te$_{0.58}$ Tamai {\it et al.} \cite{tamai} observed m*/m=20 for 
electron like band at the M point whereas it is just 6 for one of the hole like 
band at the $\Gamma$ point. It is clear from Fig (5a) and (5c) that initially 
when U ans J are not incorporated all the three hole like bands crosses the 
E$_f$ in both the compounds as well as electron like band is placed at higher 
BE (-0.29 eV) in FeSe and (-0.23 and -0.47 eV) in FeTe at the M point. Only When 
correlation effect taken into account, 
one of the hole like band comes down below the E$_f$ and other two crosses 
the E$_f$ as well as electron like band approaches towards E$_f$ as clear 
from Fig 6 where in case of FeTe  at U = 3.5 and J = 0.8 eV, 
innermost hole like band appear at 0.18 eV below E$_f$ at the $\Gamma$ point and 
electron like band position closer to E$_f$  by 0.1 eV in comparison to U and J = 0.0 eV
case. This trend qualitatively matches with above mentioned experimental findings, although 
there is a difference in the absolute energy scale. The orbital character of these near E$_f$ bands,
revealed by photoemission studies  using polarized light source, 
 \cite{chen,Miao}  also agree with our results. These results signify the importance of 
electronic correlation in these compounds however further calculation is 
required to handle the  correlation effects in a better way to bring down 
the difference between the experimental findings and calculated results based 
on LDA+U scheme.\par
From the transfer integral values, calculated by Miyake {\it et al.} for FeSe and
FeTe \cite{takashi}, it is clear that this value is large between d$_{xy}$ and 
nearest neighbour (nn) d$_{3z^2-r^2}$ orbital. It builds a strong 
interorbital hybridization between d$_{xy}$ and d$_{3z^2-r^2}$ orbitals which 
leads to a localization of d$_{3z^2-r^2}$ states at higher BE with a clear 
gap from E$_f$. The d$_{xy}$ orbitals point towards the nn Fe site so it has 
a largest band width and shows two well separated peaks one in the valence 
band  and the other in the conduction band. On the other 
hand, transfer integrals between d$_{xy}$ and nn d$_{x^2-y^2}$ orbital is 
small but for second nn it depends on the height of anion. It is large in 
case of FeSe in comparison to FeTe. Thus chalcogen-p enhance the interorbital 
hybridization between d$_{xy}$ and d$_{x^2-y^2}$ orbital which reflects in a 
prominent pseudogap structure in d$_{x^2-y^2}$ states in FeSe. Moreover, 
interorbital hybridization between d$_{xy}$ and second nn d$_{yz/xz}$ orbital is 
also mediated via anion-p orbital, hence it is also large in case of FeSe and 
contribute to pseudogap feature. On the contrary, larger height of Te anion 
allows a finite value of transfer integrals between d$_{yz/xz}$ with nn 
d$_{3z^2-r^2}$ and d$_{x^2-y^2}$ orbitals by breaking the mirror plane 
symmetry in case of FeTe. These inter orbital hybridizations are responsible 
for the gap in d$_{yz/xz}$ states at -0.8 eV, which is absent in FeSe 
(Fig(4)). When Coulomb interaction is introduced, it reduces the interorbital 
hopping and mainly d$_{yz/xz}$ and d$_{x^2-y^2}$ states are affected which 
have large number of states near E$_f$. As a consequence of this, the 
electron is transferred from the in-plane d$_{x^2-y^2}$ and d$_{xy}$ to the 
out of plane d$_{yz/xz}$ and d$_{3z^2-r^2}$ orbitals and localized them at 
higher BE in case of FeSe. In FeTe, the transfer occurs from d$_{xy}$ and 
d$_{yz/xz}$ to d$_{x^2-y^2}$ orbitals as is clear from Fig 2 and Fig 4. On 
the other hand, the application of J blocks the fluctuations in the occupancy 
of different Fe-3d orbitals which is clearly seen from 
the occupation table where a small value of J = 0.1 eV redistribute the 
electrons among different d-orbitals. A clear orbital selective effects 
(increase in the occupation of d$_{yz/xz}$ and d$_{x^2-y^2}$) is seen under 
the influence of Hund's coupling. Since crystal field splitting is large in 
FeTe because the value of $\alpha$ deviates largely from the ideal tetrahedron 
value unlike the FeSe \cite{Yin}. It increases the energy difference between 
the d$_{x^2-y^2}$ with the d$_{yz/xz}$ and d$_{xy}$ orbitals. So Hund's 
coupling promotes a gradual transfer of electrons from the highly occupied 
d$_{x^2-y^2}$ (at J = 0.0 eV) to d$_{xz/yz}$ and d$_{xy}$ orbitals with an 
increase of J value contrary to FeSe, where a small value of J is sufficient 
to transfer the electrons among these orbitals in order to reduce their 
Coulomb repulsion energy.  Thus the different strength of interorbital 
hybridization and crystal field splitting, which is mainly governed by the 
anion height, change the occupancy of electrons and band structure of 
individual Fe-3d orbital.  Hund's coupling promotes this 
differentiation and act like a band decoupler which was previously studied by 
Medici {\it et al.} \cite{Medici,Nicola}. This could be the origin of the 
orbital selective correlation effects seen in iron chalcogenide compounds.
\section{Conclusion}

We presented a systematic study of the effect of Coulomb interaction and 
Hund's coupling in Fe-3d states in FeSe and FeTe. In both the compounds 
states around E$_f$ are predominantly originated from Fe-3d orbital having a 
pseudogap feature just above the E$_f$, whereas hybridized states between 
Fe-3d and chalcogen-p orbitals lie at higher BE. This hybridization crucially 
depends on the chalcogen height from the Fe plane and it is weak in case of 
FeTe where the height of Te anion is higher in comparison to Se anion height 
in FeSe. Coulomb interaction localizes and shifts the Fe-3d states towards 
higher BE energy in both the compounds, however this interaction is strongly 
screened by Te-p bands in FeTe. It is observed that this effect is significant 
in d$_{yz/xz}$ and d$_{x^2-y^2}$ states in case of FeSe. Electrons in these 
localized states again become itinerant under the influence of J and a clear 
orbital selective changes are seen in the electronic structure. Similar to U, 
Hund's coupling effect is also prominent in FeSe in comparison to FeTe. The 
orbital selective nature of the correlation effect is linked to the different 
values of the interorbital hybridization among different Fe-d orbitals which 
is mediated via chalcogen-p orbitals. The strength of these interorbital 
hybridization mainly governs by the geometry of anion tetrahedra, height of 
anion from the Fe plane (z) and anion-Fe-anion angle {$\alpha$}. The 
difference in the anion tetrahedra geometry turns out to a different orbital 
selective nature of the correlation effect in both the compounds.

\newpage
\begin{table}
\caption{\label{table1} In this table we have presented the change in the 
occupancy of electrons in different d orbitals of Iron under the application 
of different values of U and J in FeSe and FeTe compound.}
\begin{tabular}{cccccc}
\hline
\hline
\multicolumn{6}{c}{FeSe} \\
\hline
U(eV)&J(eV) & d$_{xy}$ & d$_{xz/yz}$ & d$_{3z^2-r^2}$ & d$_{x^2-y^2}$ \\
\hline
0.0&0.0&1.239&1.300&1.461&1.363\\
4.0 &0.0& 0.954 & 1.581 & 1.655 & 0.792\\
4.0&0.1&1.073&1.257&1.765&1.436\\
4.0&1.2&1.189&1.203&1.593&1.526\\
\hline
\hline
\multicolumn{6}{c}{FeTe} \\
\hline
0.0&0.0&1.242&1.308&1.462&1.363\\
3.5&0.0&1.182&1.109&1.569&1.637\\
3.5&0.1&1.174&1.138&1.558&1.634\\
3.5&1.2&1.189&1.209&1.514&1.572\\
\hline
\hline
\end{tabular}
\end{table}       
\newpage
\begin{figure}
\includegraphics[width=12cm,keepaspectratio]{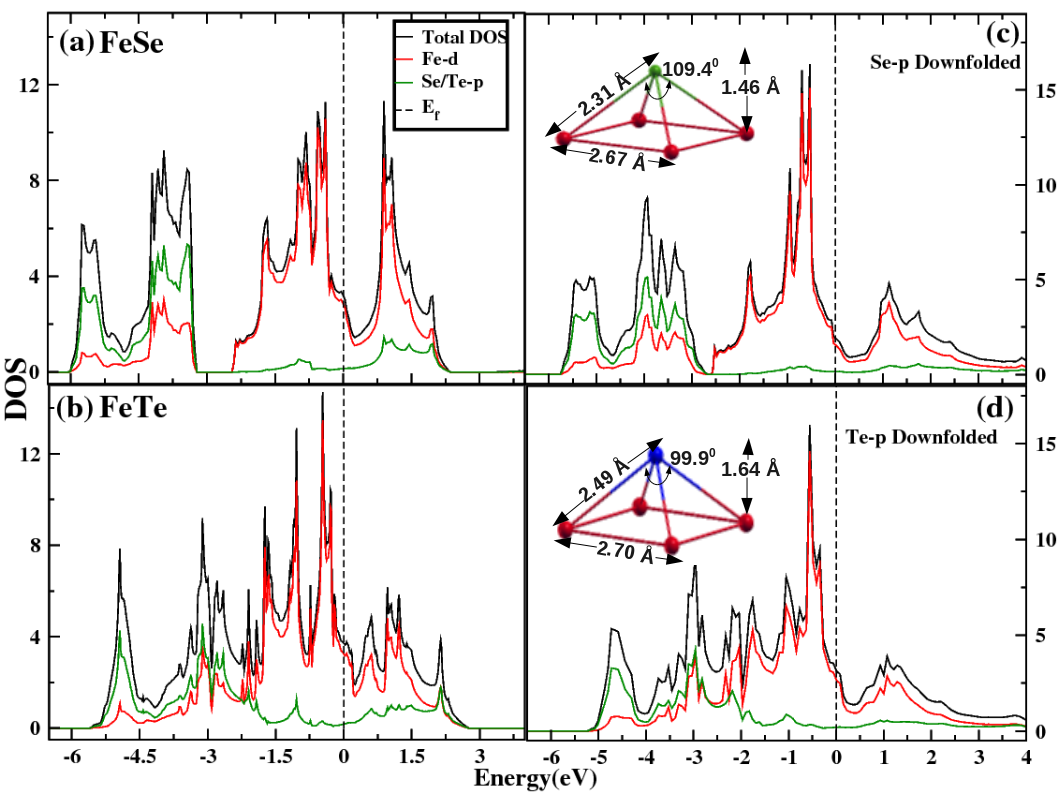}
\caption{\label{ldoscom} Total and partial DOS have been plotted for FeSe in 
(a) and for FeTe in (1b). In (c) and (d) total and partial DOS have been 
calculated, where Se-p and Te-p valence orbitals are doenfolded in FeSe and 
FeTe respectively. Inset of (c) and (d) shows the anion tetrahedra in FeSe 
and FeTe respectively. Fe, Se and Te atoms are denoted by red, blue and green 
clours respectively in the inset pictures (c) and (d).}
\end{figure}
\newpage
\begin{figure}
\includegraphics[width=6cm,keepaspectratio]{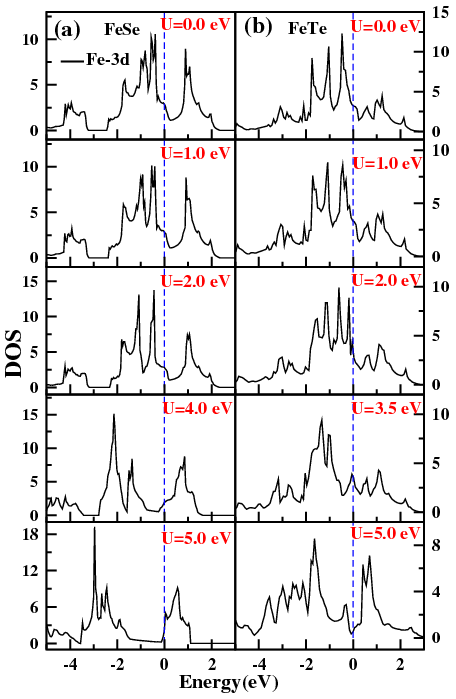}
\caption{\label{ldoscom} Panel (a) shows changes in Fe-3d states in FeSe at 
different values of U, where the value of U is written in red colour in each 
plot. Panel (b) shows changes in Fe-3d states in FeTe at different values of 
U, where the value of U is written in red colour in each plot.}
\end{figure}
\newpage
\begin{figure}
\includegraphics[width=6cm,keepaspectratio]{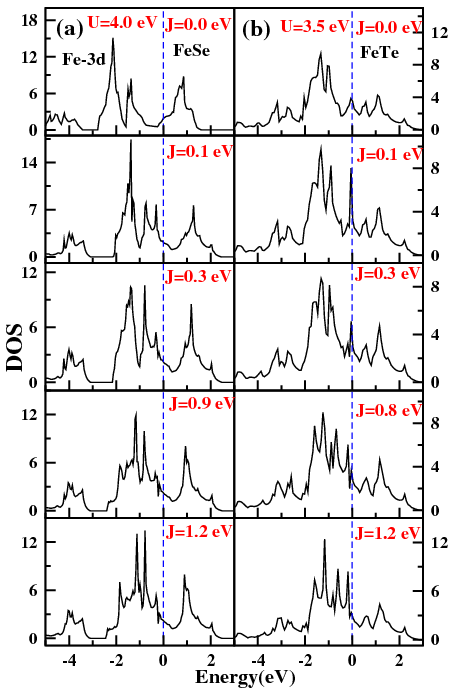}
\caption{\label{ldoscom}  Panel (a) shows changes in Fe-3d states in FeSe at 
different values of J, where the value of J is written in red colour in each 
plot. Panel (b) shows changes in Fe-3d states in FeTe at different values of 
J, where the value of J is written in red colour in each plot. The value of U 
is  4.0 and 3.5 eV in all FeSe and FeTe plots respectively.}
\end{figure}
\newpage
\begin{figure}
\includegraphics[width=10cm,keepaspectratio]{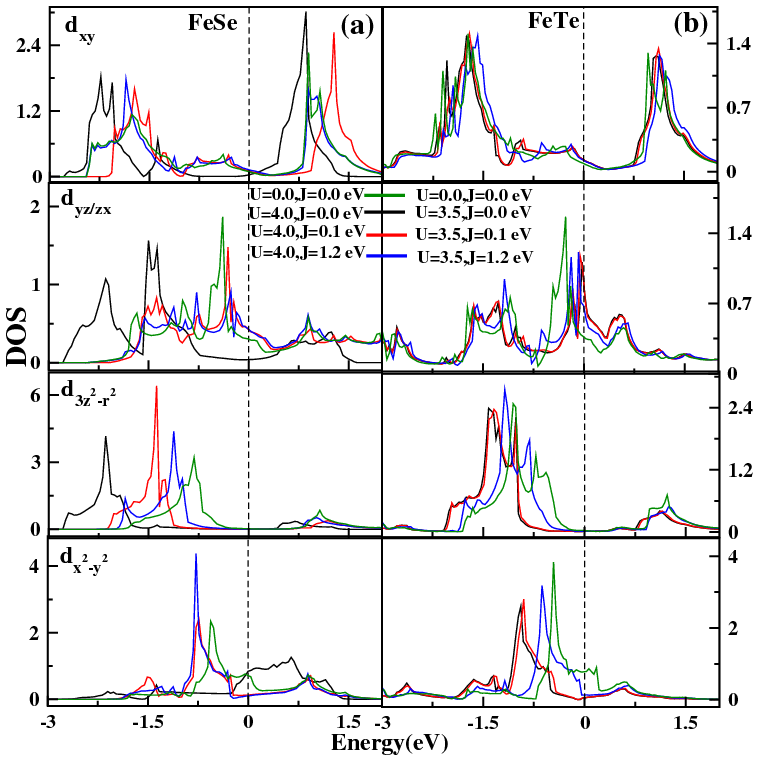}
\caption{\label{mdoscom}Panel (a) shows changes in the DOS of different d 
orbitals of Fe atom in FeSe at different values of U and J. Panel (B) shows 
changes in the DOS of different d orbitals of Fe atom in FeTe at different 
values of U and J. Different values of U and J are denoted by Green, Black, 
Red and Blue colours at the top of the Figure.}
\end{figure}
\newpage
\begin{figure}
\includegraphics[width=12cm,keepaspectratio]{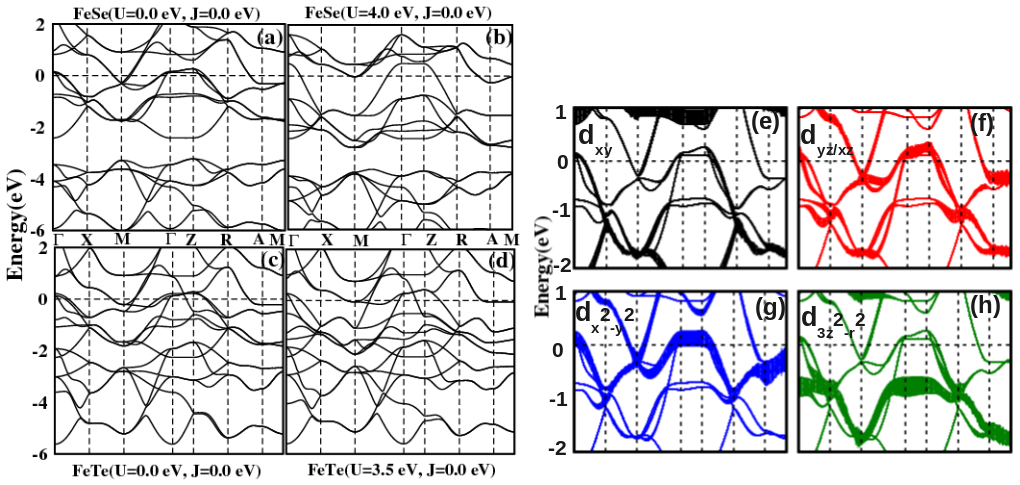}
\caption{\label{ldoscom} (a) and (b) shows band structure plot of FeSe at U = 
0.0 and U = 4.0 eV respectively. (c) and (d) shows band structure plot of 
FeTe at U = 0.0 and U = 3.5 eV respectively. (e), (f), (g) and (h) shows 
fatness of bands which are originated from d$_{xy}$, d$_{yz/xz}$, 
d$_{x^2-y^2}$ and d$_{3z^2-r^2}$ respectively for FeSe at U and J = 0.0 eV.}
\end{figure}
\newpage
\begin{figure}
\includegraphics[width=12cm,keepaspectratio]{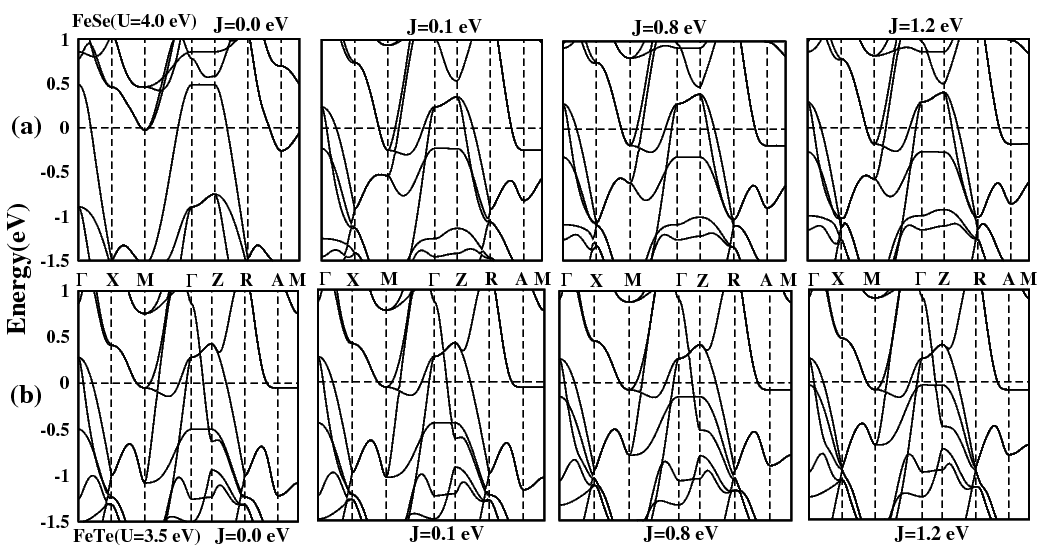}
\caption{\label{ldoscom} Panel (a) and (b) shows  band structure plot for FeSe
and FeTe at different values of J = 0.0, 0.1, 0.8 and 1.2 eV from left to 
right respectively. The value of U is 4.0 and 3.5 eV in all FeSe and FeTe 
plots respectively.}
\end{figure} 
\end{document}